\documentclass[12pt,preprint]{emulateapj}

\def\be{\begin{equation}}
\def\ee{\end{equation}}
\def\bea{\begin{eqnarray}}
\def\eea{\end{eqnarray}}

\usepackage{graphicx}

\usepackage{xcolor}
\usepackage[normalem]{ulem}

\begin{document}

\title{Revealing physical activity of GRB central engine with macronova/kilonova data}
\author{Zhao-Qiang Shen$^{1,2}$, Zhi-Ping Jin$^{1}$, Yun-Feng Liang$^{1,2}$, Xiang Li$^{1}$, Yi-Zhong Fan$^{1}$, and Da-Ming Wei$^{1}$}
\affil{
$^1$ {Key Laboratory of dark Matter and Space Astronomy, Purple Mountain Observatory, Chinese Academy of Science, Nanjing, 210008, China.}\\
$^2$ {University of Chinese Academy of Sciences, Yuquan Road 19, Beijing, 100049, China.}
}
\email{yzfan@pmo.ac.cn (YZF) and dmwei@pmo.ac.cn (DMW)}

\begin{abstract}
The modeling of Li-Paczy\'{n}ski macronova/kilonova signals gives reasonable estimate on the neutron-rich material ejected during the neutron star mergers.
%in three short or hybrid Gamma-ray Bursts (GRBs) is strongly in support of the neutron star merger origin.
Usually the accretion disk is more massive than the macronova ejecta,
with which the efficiencies of converting the disk mass into prompt emission of three merger-driven GRBs can hence be directly constrained.
Supposing the { macronovae}/kilonovae associated with GRB 050709, GRB 060614 and GRB 130603B arose from radioactive decay of the r-process material, the { upper limit on} energy conversion efficiencies are found to be as low as $\sim 10^{-6}-10^{-4}$.
Moreover, for all three events, neutrino annihilation is likely powerful enough to account for the brief gamma-ray flashes. Neutrino annihilation can also explain the ``extended" emission lasting $\sim 100$ s in GRB 050709, but { does} not work for the one in GRB 060614.
These progresses demonstrate that the macronova can serve as a novel probe of the central engine activity.
\end{abstract}

\section{Introduction}
The researches on mergers of a neutron star with either a neutron star or a stellar-mass black hole have attracted wider and wider attention since they are promising sources of significant gravitational wave (GW) radiation that are expected to be detectable for the advanced detectors such as LIGO and Virgo in the near future \cite{Abbott2016}. Currently, the ``related" observational data are mainly for short Gamma-ray Bursts (sGRBs; the bursts with a duration shorter than 2 sec), a kind of brief $\gamma-$ray transient widely believed to originate from compact object mergers \citep{Eichler1989}, as favored by a broad range of observations \citep{Berger2014}. With the sGRB data, our understanding of the rate and the launched ultra-relativistic outflow (including the subsequent radiation) of neutron star mergers have been significantly advanced \citep{Nakar2007,Berger2014}. The knowledge of the central engine, such as the accretion disk mass, the spin and the mass of the newly formed black hole, and the energy extraction efficiency, however, is almost solely from the numerical simulations \citep{Ruffert1998,Rosswog1999,Kiuchi2009,Rezzolla2011} because the detectable photons were emitted at radii far from the central engine and usually the initial information of the GRB ejecta was lost. One exception is for the GRBs with a distinct thermal radiation component the standard fireball acceleration and radiation apply \citep{Piran1992,Meszaros1993}, for which the initial radius at which the fireball was launched or reformed can be reasonably inferred and a constraint on the mass of the central black hole is imposed \citep{Peer2007,Fan2011}. However, no distinct thermal radiation component has been identified in prompt emission of sGRBs, yet.

Thanks to the relatively narrow mass distribution of the double neutron star binaries observed in the Galaxy \citep{Lattimer2012}, the mass of the nascent black hole ($M_{\rm BH}$)
and its spin parameter ($a$) can be reliably reasonably evaluated, as demonstrated in both analytical approaches and numerical simulations \citep{Lee2000,Kiuchi2009}. Adopting these prior values of $M_{\rm BH}$ and $a$, with the electromagnetic data the mass of the accretion disk ($M_{\rm disk}$) launching the sGRB ejecta can be estimated \citep{Fan2011}. Not surprisingly, the estimated accretion disk masses are sensitively dependent of the energy extraction processes since the magnetic and neutrino mechanisms have rather different energy extraction efficiency  \citep{Fan2005}. As a result, there are large divergency between the accretion disk masses inferred in different energy extraction models \citep{Fan2011,Giacomazzo2013,Liu2015} and suffer from further uncertainties involved in the modeling. The other hope is that for some ``nearby" ($\sim 100$ Mpc) binary-neutron-star mergers, with
the gravitational wave data the masses of the binaries as well as the formed accretion disk might be inferable in 2020s \citep{Kiuchi2010}. A short summary of the current situation is the lack of solid information about the central engine directly inferred from the observational data of sGRBs.

The neutron-rich sub-relativistic matter ejected in the neutron star mergers are the ideal sites to synthesize the r$-$process material \citep{Lattimer1974,Eichler1989}.
The radioactive decay of these heavy elements gives rise to the near-infrared/optical transients, i.e., the so-called Li-Paczy\'{n}ski macronova/kilonova \citep{Li1998,Metzger2010,Kasen2013}.
Recently macronova/kilonova signals have been identified
%, the optical/near-infrared transient arising from radioactive decay of r$-$process material synthesized in ejecta launched during the mergers \citep{Lattimer1974,Eichler1989},
in a few events \citep{Tanvir2013,Berger2013,Yang2015,Jin2015,Jin2016}. Such new transients, in addition to labeling the birth sites of the very heavy elements, are promising electromagnetic counterparts of gravitational sources that are expected to be detectable by advanced LIGO/Virgo in the near future. In this work we show that {\it the macronova observations also serves as a novel probe of the central engine activity of sGRBs.}

\section{The method}
Throughout this work we assume that the { macronovae}/kilonovae associated with GRBs arose from the radioactive decay of the r-process material launched by neutron star mergers, which is particularly motivated by the fact that a NS$-$BH merger model can reasonably reproduce the multi-epoch/band light curve of the macronova associated with GRB 060614 \citep[see Fig.1 of][]{Jin2015}. An independent/important support to the scenario of generating $\sim 0.1~M_\odot$ r-process material in a neutron star merger event is the heavy element enrichment detection of the ancient dwarf galaxy Reticulum II \citep{Ji2016}. Alternatively the current GRB-associated macronova/kilonova signals could be for example the thermal-like radiation of the sub-relativistic ejecta heated by the central engine-generated X-rays \citep{Kisaka2016}. In such a model the sub-relativistic ejecta mass can be much less than that found in the r-process material modeling and the progenitor stars may be double NSs. Nevertheless, as long as the ejecta mass is found in a given macronova/kilonova origin model, the approaches outlined in this work can be straightforwardly applied to. Hence in this work we focus on our ``fiducial" case.

Within the r-process material radioactive decay scenario, the modeling of the macronova/kilonova lightcurve can yield a reasonable estimate of the mass of the neutron-rich ejecta ($M_{\rm ej}$) and possibly also distinguish between the progenitor stars, either binary neutron stars or neutron star-black hole binary. The major differences of the ejecta from these two kinds of progenitors include \citep{Hotokezaka2013,Tanaka2014}:
a) some NS$-$BH mergers eject much more material than the NS binary mergers;
b) %\szqd{the NS$-$BH merger ejecta is more collimated than the NS binary merger ejecta.}
NS-BH merger ejecta are launched predominantly along the orbital plane.
Consequently, the macronova powered by some NS$-$BH mergers are much more luminous, bluer and lasts longer than the macronovae powered by NS binary mergers.
Such a goal likely has already been achieved in the studies of some GRB/macronova events and NS$-$BH merger origin is favored
\citep{Hotokezaka2013,Yang2015,Jin2016,Kawaguchi2016}.
Below we focus on probing the central engine activity with $M_{\rm ej}$ within the NS$-$BH merger scenario.

\begin{figure}[!h]
\includegraphics[width=0.5\textwidth]{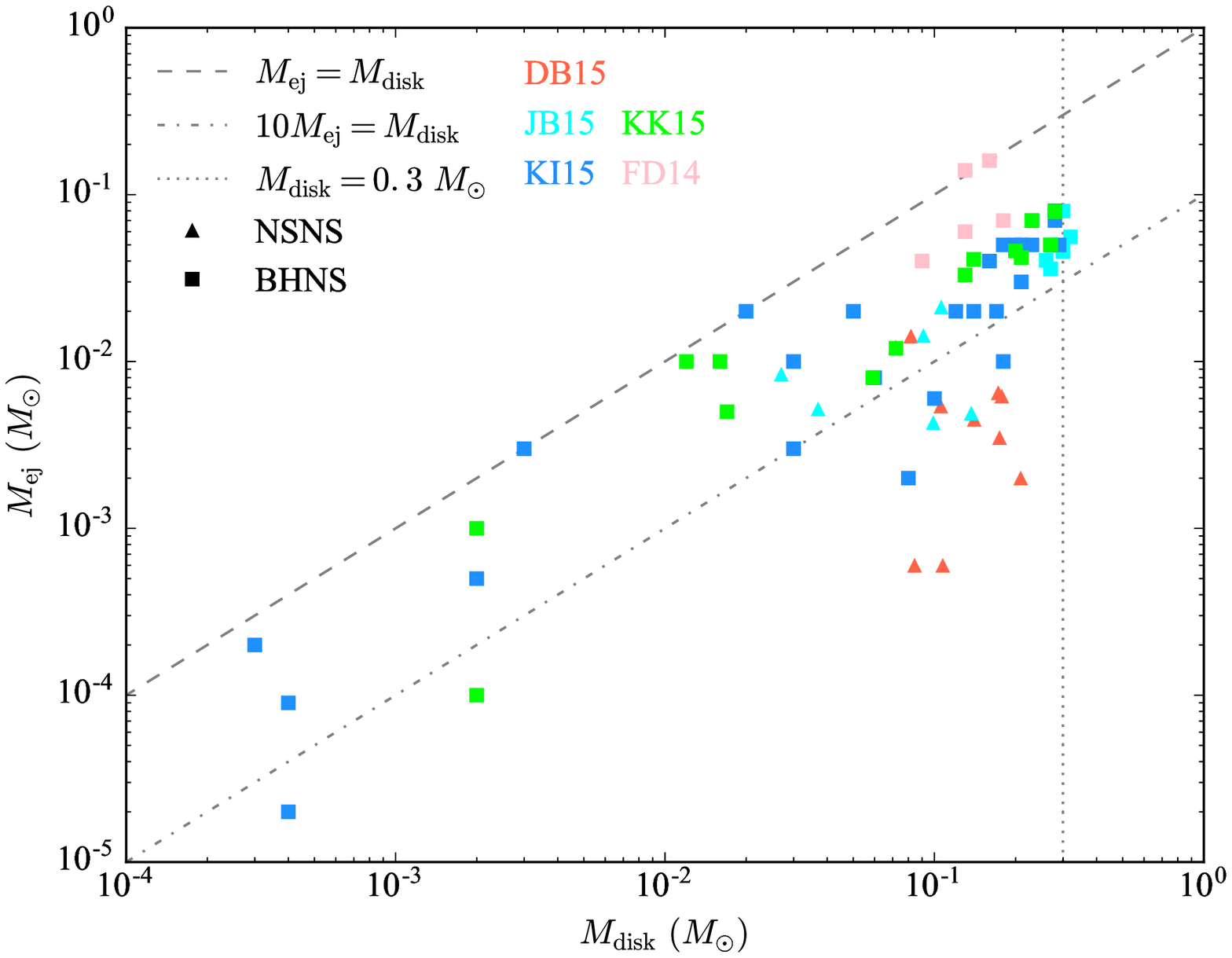}
\caption{
The remnant disk rest masses and the ejecta masses after binary compact object mergers in two latest numerical relativity simulations.
Triangles and squares represent the simulations of binary neutron star (NSNS) mergers reported in Dietrich et al. \citep{Dietrich2015} and the simulations of black hole-neutron star (BHNS) mergers reported in \citep{Kawaguchi2015,Just2015,Kyutoku2015,Foucart2014}, respectively. The dashed line is for $M_{\rm disk}=M_{\rm ej}$. A clear/general relationship is $M_{\rm disk}\geq M_{\rm ej}$ and there is likely also an upper limit of $M_{\rm disk}\leq 0.3~M_\odot$.}
\label{Mdisk_Mej}
\end{figure}

The central engine activity is ``governed" by $M_{\rm disk}$. Previously the $M_{\rm disk}$ can only be inferred with the electromagnetic data within the double neutron star merger scenario \citep{Fan2011,Giacomazzo2013,Liu2015}, which is strongly-model dependent. Such a puzzle, fortunately, can be solved in a new way.  As found in the advanced numerical simulations on neutron star mergers (including both the double neutron star mergers and the neutron star$-$stellar-mass black hole mergers; see \citep{Foucart2014,Dietrich2015,Kawaguchi2015,Just2015,Kyutoku2015} and the references therein), there is a general conclusion that ``the accretion disk mass $M_{\rm disk}$ is (significantly) larger than $M_{\rm ej}$" (see Fig.\ref{Mdisk_Mej} for illustration; where the triangles represent the mergers of equal- and unequal-mass neutron stars assuming several different equations of state (EOS), while the squares are the black hole-neutron star mergers with various initial parameters). One exception is in the case of tidal disruption of a star with a hyperbolic orbit, half the material is ejected and another half is bound (i.e., $M_{\rm disk}=M_{\rm ej}$; Hotokezaka 2016 private communication). Therefore we have $M_{\rm disk}\geq M_{\rm ej}$ if a black hole was promptly formed in the mergers. For the neutron star$-$stellar-mass black hole mergers it is always the case. If instead a hypermassive neutron star was formed in the binary neutron star mergers, it will take a time $\sim 100$ ms or longer to collapse into a black hole, during which most of the disk material likely has accreted onto the central remnant and $M_{\rm ej}$ can not be taken as a robust lower limit of $M_{\rm disk}$. That is why we focus on the NS$-$BH mergers favored in current macronova/kilonova modeling. According to Fig.1 it is also reasonable to take $M_{\rm disk}\leq 0.3~M_\odot$.

%The main goal of this work is to propose a straightforward but novel approach to reasonably estimate
The energy extraction efficiency ($\epsilon$) of the sGRB central engine  is
\begin{equation}
\epsilon =E_{\rm \gamma,j}/M_{\rm disk} c^{2},
\label{eq_efficiency}
\end{equation}
where $E_{\rm \gamma,j}=E_\gamma (1-\cos \theta_{\rm j})\approx E_\gamma \theta_{\rm j}^{2}/2$ is the geometry-corrected prompt emission energy of the sGRB, $\theta_{\rm j}\ll 1$ is the half-opening angle of the GRB ejecta and $E_\gamma$ is the isotropic-equivalent energy of the prompt emission. With the good-quality prompt and afterglow emission data of some sGRBs, $E_\gamma$ can be directly measured and the $\theta_{\rm j}$ can be yielded in the numerical modeling of the afterglow emission, with which $E_{\rm \gamma,j}$ is obtained. The main challenge is how to measure or estimate $M_{\rm disk}$ reliably. Fortunately we have shown that $M_{\rm disk}\geq M_{\rm ej}$, with which $\epsilon$ can be constrained.

 Thanks to the great efforts of the GRB follow-up observation community, three macronova signals have been identified in sGRB 130603B \citep{Tanvir2013,Berger2013}, hGRB 060614 (hybrid GRB; \citep{Yang2015,Jin2015}) and sGRB 050709 \citep{Jin2016}. The numerical modeling of the signal in sGRB 130603B in the binary neutron star merger scenario suggests a $M_{\rm ej}\sim 0.03-0.08~M_\odot$ \citep{Berger2013}, which might be a challenge since usually the binary neutron star mergers are not expected to be able to eject such massive outflow. On the other hand, a neutron star-black hole merger model \citep{Tanaka2014} can reasonably reproduce the macronova signal in sGRB 130603B with a $M_{\rm ej}\sim 0.05~M_\odot$ \citep{Hotokezaka2013,Kawaguchi2016,Jin2016}. For hGRB 060614, the macronova modeling gives a $M_{\rm ej} \sim 0.1~M_\odot$ in the neutron star-black hole merger scenario and $\sim 0.2~M_\odot$ in the case of binary neutron star merger \citep{Yang2015}. The $I$/F814W-band macronova signal of sGRB 050709 is rather similar to (though a bit dimmer than) that of hGRB 060614 and the neutron star-black hole merger modeling suggests a $M_{\rm ej}\sim 0.05~M_\odot$ \citep{Jin2016}. Other physical parameters of these three macronova-associated GRBs, including $E_\gamma$ and $\theta_{\rm j}$, are also summarized in Tab.1. With eq.(\ref{eq_efficiency}) and the fact that $M_{\rm disk}\geq M_{\rm ej}$, it is thus straightforward to get the constraint on $\epsilon$, which is found to be very low.

%%------------
\begin{table*}
\begin{center}
\caption{The estimate of the energy extraction efficiency}
\label{tab-GRBs}
\begin{tabular}{lccc}
\hline
							& GRB 050709$^{a}$				& GRB 060614$^{b}$				& GRB 130603B$^{c}$		\\
\hline
$E_{\rm \gamma}$ ($10^{51}$ erg)	& 0.069 			& 2.1							& 2.1 					\\
%$z$							& 0.16							& 0.125							& 0.356					\\
Duration (sec)$^{d}$	& 0.07(+130) 			&  5(+97) 							&            0.18 					\\
$\theta_{\rm j}$ (rad)			& 0.10 							& 0.08							& 0.085 				\\
$M_{\rm ej}~(M_\odot)^{e}$ 	& $\sim 0.05$					& $\sim 0.10$					& $\sim 0.05$			\\
\hline
$M_{\rm disk}~(M_\odot)$ 	& $\gtrsim 0.05$				& $\gtrsim 0.10$				& $\gtrsim 0.05$			\\
$\epsilon^{f}$					& $\lesssim 3.9 \times 10^{-6}$	& $\lesssim 3.8 \times 10^{-5}$ & $\lesssim 8.4 \times 10^{-5}$	\\
$L_{\nu\bar{\nu}}$ for initial spike					& Yes	& Possible & Possible	\\
$L_{\nu\bar{\nu}}$ for extended emission				& Possible	& No & 	\\
%$\epsilon_{\rm jet}$$^{d}$	& $\lesssim 3.9 \times 10^{-5}$	& $\lesssim 4.5 \times 10^{-4}$ & $\lesssim 1.4 \times 10^{-3}$	\\
\hline
\end{tabular}
\end{center}
Note:
a. \citet{Villasenor2005} and \citet{Jin2016};\\
b. \citet{Gehrels2006} and \citet{Xu2009};\\
c. \citet{Tanvir2013}, \citet{Berger2013} and \citet{Fan2013};\\
d. The durations include that of the hard spike and the `extended emission' (in the bracket);\\
e. The value of $M_{\rm ej}$ is estimated within the neutron star-black hole merger model;\\
f. The value of $\epsilon$ is derived with eq.(\ref{eq_efficiency}).
\end{table*}
%%-------------

Let us check either the neutrino/anti-neutrino annihilation process or the magnetic process is needed to account for the data. Note that in typical neutron star-black hole mergers, an initial spin $a>0.6$ and a $M_{\rm BH}<10M_\odot$ (the observed Galactic black holes likely have a typical mass $\sim 6-8~M_\odot$) are needed to eject massive neutron-rich outflow. Therefore in the following discussion we take $a=0.7$ and $M_{\rm BH}=7M_\odot$. An empirical relation of the luminosity of such a hot ejecta reads \citep{Zalamea2011}
\begin{eqnarray}
L_{\nu\bar{\nu}} &\approx & 5\times 10^{52}~{\rm erg~s^{-1}}~({\theta_{\rm j}\over 0.1})^{-2}({x_{\rm ms}\over 1.7})^{-4.8}({\dot{M}\over 1~M_\odot~{\rm s}^{-1}})^{9/4}\nonumber\\
&&({M_{\rm BH}\over 7M_\odot})^{-3/2},
\label{eq:Lvv}
\end{eqnarray}
which holds for $\dot{M}_{\rm ign}<\dot{M}<\dot{M}_{\rm trap}$, where $x_{\rm ms}=\{3+Z_{2}\mp [(3-Z_{1})(3+Z_{1}+2Z_{2})]^{1/2}\}/2$ (where $Z_{1}=1+(1-a^{2})^{1/3}[(1+a)^{1/3}+(1-a)^{1/3}]$ and $Z_{2}=(3a^{2}+Z_{1}^{2})^{1/2}$; For $a=(0.8,~0.9,~0.95)$ we have $x_{\rm ms}=(1.45,~1.16,~1.00)$, respectively); $\dot{M}$ is the accretion rate, $\dot{M}_{\rm ign}=K_{\rm ign}(\alpha/0.1)^{5/3}$, $\dot{M}_{\rm trap}=K_{\rm trip}(\alpha/0.1)^{1/3}$ and $\alpha$ is the viscosity. Both $K_{\rm ign}$ and $K_{\rm trap}$ are functions of $a$ and for $a=(0,~0.95)$, $K_{\rm ign}=(0.071,~0.021)~M_\odot~{\rm s}^{-1}$ and $K_{\rm trap}=(9.3,~1.8)~M_\odot~{\rm s}^{-1}$, respectively \citep{Zalamea2011}.
The corresponding efficiency of ``energy extraction" is
\begin{eqnarray}
\epsilon_{\nu\bar{\nu}} &\equiv & {\theta_{\rm j}^{2}\over 2}{L_{\nu\bar{\nu}}\over \dot{M}c^{2}} \nonumber\\
&\approx & 7\times 10^{-6}~({x_{\rm ms}\over 1.7})^{-4.8}({\dot{M}\over 0.1~M_\odot~{\rm s}^{-1}})^{5/4}({M_{\rm BH}\over 7M_\odot})^{-3/2}.
\label{eq:nunu}
\end{eqnarray}
The smaller the $\dot{M}$ is, the less efficient the energy extraction we have.

In the Blandford$-$Znajek mechanism \citep{Blandford1977,Lee2000}, the
luminosity of the electromagnetic outflow can be estimated by
\begin{eqnarray}
L_{\rm BZ} &\approx & 1.5\times 10^{52}~{\rm erg~s^{-1}}~\varepsilon_{\rm B}({a\over 0.7})^{2}({\dot{M}\over 0.01~M_\odot~{\rm s}^{-1}})({\theta_{\rm j}\over 0.1})^{-2}\nonumber\\
&& [(1+\sqrt{1-a^{2}})/2]^{-2},
\label{eq:Lbz}
\end{eqnarray}
%$R_{\rm H}=(1+\sqrt{1-a^{2}})r_{\rm g}/2$, $r_{\rm g}=2GM_{\rm BH}/c^2\approx 20~{\rm km}~(M_{\rm BH}/7M_\odot)$ is the Schwarzschild radius and $G$ is the gravitational constant. \red{Need to be re-arranged into a better version!}
%估算磁场是比较复杂，我们组一般用的是冲压平衡关系：
%By Weihua Lei: $B^2 /(8\pi) = \rho c^2 = \dot{M}c/(4\pi r_H^2)$,
%where $r_H=(1+\sqrt{1-a^2}) GM/c^2$ is the radius of the black hole horizon.
where
$\varepsilon_{\rm B}\sim {\cal O}(1)$ is a dimensionless parameter to describe the ratio between the ordered magnetic field energy density and the total energy density of the accreting material.
The energy extraction efficiency is thus
\begin{equation}
\epsilon_{\rm BZ}\equiv {\theta_{\rm j}^2\over 2}{L_{\rm BZ}\over \dot{M}c^{2}}\approx 3.8\times 10^{-4}~{\varepsilon_{\rm B}\over 0.1}({a\over 0.7})^{2}[(1+\sqrt{1-a^{2}})/2]^{-2}.
\label{eq:e_BZ}
\end{equation}
In the case of $\dot{M}\ll 1~M_\odot~{\rm s^{-1}}$ we have $\epsilon_{\nu\bar{\nu}} \ll \epsilon_{\rm BZ}$ (see eq.(\ref{eq:nunu}) and eq.(\ref{eq:e_BZ})).

\section{Case studies}

Below let us discuss these three GRBs and their ``extended emission" case by case. In particular we focus on whether the neutrino process can work or not since all the $\epsilon$ reported in Tab.1 are sufficient small to be well matched (see eq.(\ref{eq:e_BZ})). To be able to account for the data within the neutrino process the requests of $L_{\nu\bar{\nu}}\geq L_{\tau}$ (equally, $\epsilon_{\nu\bar{\nu}}\geq \epsilon_{\tau}$) and $\dot{M}\leq M_{\rm disk}/\tau$ should be satisfied, where ($\tau,~L_\tau,~\epsilon_{\tau}$) are the (duration after the redshift correction, luminosity, energy extraction efficiency) of the prompt emission or alternatively the extended emission, respectively. Note that $M_{\rm ej}\leq M_{\rm disk}\leq 0.3M_\odot$ and $\epsilon_{\tau}$ is different from $\epsilon$ in the case of presence of extended emission.

{\it GRB 050709.} It is a short burst lasting $\sim 0.07$ s followed by the extended X-ray emission with a duration of $\sim 130$ s.
At a redshift of $z=0.16$, the initial hard spike has a $E_\gamma \sim 2.8\times 10^{49}~{\rm erg}$ and a corresponding luminosity of $L_{\rm \tau=0.06~s}\sim 4.5\times 10^{50}~{\rm erg~s^{-1}}$ \citep{Villasenor2005}. With eq.(\ref{eq:Lvv}) we find out that with the other fiducial parameters, for $\dot{M}\sim 0.15~{M_\odot ~ \rm s^{-1}}$ the neutrino/anti-neutrino process is energetic enough to generate the hard spike. The corresponding accretion disk mass is $\sim 0.01~M_\odot$, significantly smaller than $M_{\rm ej}\sim 0.05~M_\odot$. Therefore most of the accretion disk mass might have been consumed to yield the  $\sim 112~(1+z)$ sec long-tail X-ray emission.  The time-averaged luminosity of the extended X-ray emission is $L_{\rm \tau=112~{\rm s}}\sim 3\times 10^{47}~{\rm erg~s^{-1}}$, the neutrino process may be able to work with a high spin parameter $a\sim 0.9$ (i.e., $x_{\rm ms}=1.16$) and $\dot{M}\sim 2\times 10^{-3}~M_\odot~{\rm s}^{-1}$. The disk mass for the extended emission is $\sim 0.22~M_\odot$, which may be possible in some neutron star$-$black hole merger models (see Fig.1). The $\sim 100$ s duration of the ``steady extended" accretion process for the disk formed in the compact object mergers requires a $\alpha \sim 10^{-3}$ (see also \citep{Lee2009}), which is at the low end of the distribution discussed in the literature (note that this request also holds for the magnetic energy extraction process). As far as the energy budget is concerned, the neutrino process seems to be plausible for GRB 050709.

{\it GRB 130603B.} The burst was at a redshift of $z=0.356$ and the prompt emission lasted for $\sim 0.12(1+z)$ s with a luminosity of $L_{\rm \tau=0.12~{\rm s}}\sim 1.8\times 10^{52}~{\rm erg~s^{-1}}$ \citep{Tanvir2013}. With eq.(\ref{eq:Lvv}) we find out that with  other adopted  fiducial parameters, for $\dot{M}\sim 0.6~{M_\odot ~ \rm s^{-1}}$ the neutrino/anti-neutrino process is energetic enough to generate the brief but intense gamma-ray flash (see also eq.(\ref{eq:nunu})).
This requires a $M_{\rm disk}\sim 0.07~M_\odot$, which seems possible with an inferred $M_{\rm ej}\sim 0.05~M_\odot$ (see Fig.1).
%the neutrino/anti-neutrino process can give rise to the short gamma-ray flash, too.

{\it GRB 060614.}
The prompt emission of hGRB 060614 at $z=0.125$ consisted of two epoches, the first is a hard spike lasting $\sim 4.4(1+z)$ sec with a $E_\gamma\sim 3.7\times 10^{50}$ erg \citep{Xu2009}, suggesting a corresponding luminosity $L_{\tau=4.4~{\rm s}}\sim 8\times 10^{49}~{\rm erg~s^{-1}}$. The neutrino/anti-neutrino process can marginally give rise to the initial gamma-ray flash with a $M_{\rm disk}\sim 0.2~M_\odot$ and $a\sim 0.8$. In view of the rather massive ejecta with $M_{\rm ej}\sim 0.1M_\odot$, a $M_{\rm disk} \sim 0.2~M_\odot$ is possible. However, for the $\sim 86~(1+z)$ sec long-tail emission of hGRB 060614 with an averaged luminosity of $L_{\rm \tau=86~s}\sim 2\times 10^{49}~{\rm erg~s^{-1}}$, the neutrino process is hard to contribute. This is because the total accretion disk is not expected to be more massive than $0.3~M_\odot$ (see Fig.1), with which the averaged accretion rate $\dot{M}<3\times10^{-3}~M_\odot~{\rm s^{-1}}$. In the most promising case (i.e., a very high $a\sim 0.95$ and a sufficient small $\alpha\leq 0.01$ to render $\dot{M}_{\rm ign}\leq 10^{-3}~M_\odot~{\rm s}^{-1}$) we have a luminosity of $L_{\nu\bar{\nu}}\sim  10^{48}~{\rm erg~s^{-1}}$, one order of magnitude lower than the observed value. On the other hand, we have $L_{\rm BZ}\approx 1.5\times 10^{51}~{\rm erg~s^{-1}}~\varepsilon_{\rm B}(a/0.7)^{2}(\dot{M}/10^{-3}~M_\odot~{\rm s^{-1}})$, which can match the observation data as long as $\varepsilon_{\rm B}>0.01~(a/0.7)^{-2}(\dot{M}/10^{-3}~M_\odot~{\rm s^{-1}})^{-1}$. The initial magnetization degree (i.e., the initial ratio between the magnetic field energy density and the thermal energy density) of the outflow powering the long-tail emission is expected to be $\sim$ tens. Again, the $\sim 100$ s duration of the accretion process for the disk formed in the compact object mergers requires a $\alpha \sim 10^{-3}$.

\begin{figure}[!h]
\includegraphics[width=0.5\textwidth]{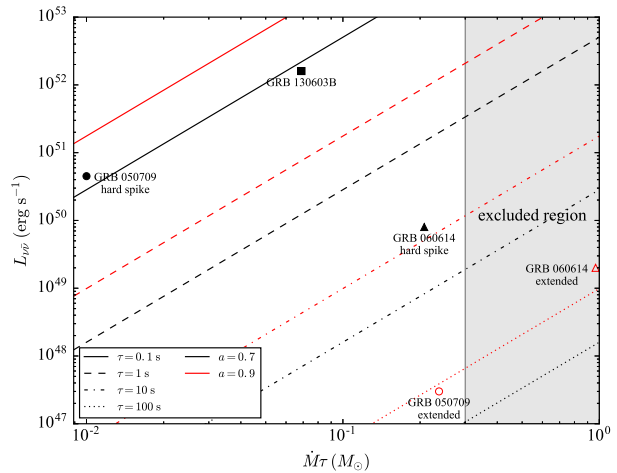}
\caption{
The neutrino anti-neutrino annihilation luminosity ($L_{\nu\bar{\nu}}$ for $\theta_{\rm j}=0.1$) as a function of $\dot{M}\tau$. The (solid, dashed, dash-dotted, dotted) lines are for $\tau=(0.1,~1.0,~10.0,~100.0)$ s, respectively. The lines and symbols in (black, red) are for $a=(0.7,~0.9)$, respectively.}
\label{fig:constraints}
\end{figure}

All these results/constraints in the case of neutrino anti-neutrino annihilation are summarized in Fig.\ref{fig:constraints}. Clearly the neutrino process may be able to account for some  but not all data and the magnetic process is in principle sufficient for all the data.

If $M_{\rm ej}\sim 10^{-3}-10^{-2}M_\odot$, as suggested in \citet{Kisaka2016}, the limits on the efficiencies of the central engines will be 1-2 orders of magnitude higher,
{ leading to  looser constraints}. For $M_{\rm BH}\sim 3M_\odot$ (i.e., within the double neutron star merger scenario), with a reasonable $a\sim 0.7$ the accretion disk masses $M_{\rm disk}$ are required to be $\sim (0.005,~0.04,~0.12)~M_{\odot}$ for the hard spikes of GRB 050709, GRB 060614 and GRB 130603B respectively, otherwise the neutrino process is not efficient enough. { Though in some binary neutron star mergers $M_{\rm disk}$ can be as massive as $\sim 0.2~M_\odot$ (see the triangles in Fig.1), the inferred rather-low $M_{\rm ej}$ hampers us to draw a reliable conclusion on the role of the neutrino process.}
For the extended X-ray emission of GRB 050709 and GRB 060614, the required $M_{\rm disk}\sim (0.38,~1.6)~M_\odot$ even for a somewhat ``optimistic” $a=0.8$. We thus suggest that
the neutrino process is hard to account for the extended emission.

\section{Conclusion and discussion}
We show in this work the macronova data provide a novel probe of the GRB central engine activity. The conversion efficiency of the disk mass into the GRB prompt emission is found to be rather low and the neutrino process are likely able to generate the brief gamma-ray flashes in all three events (see Tab.1 and Fig.2). The situation is less clear for the extended emission found in GRB 050709 and GRB 060614. For the former, the neutrino process may work thanks to the low luminosity of the extended X-ray emission. While for the latter, the magnetic process is necessary. Note that these conclusions are based on the NS$-$BH merger model for current GRB-associated { macronova}/kilonova events. Interestingly, all current three macronova-associated GRBs have small offsets from the host Galaxy centers, in agreement with the NS$-$BH merger model prediction \citep{Troja2008}. { In other scenarios that yield a much lower $M_{\rm ej}$, the constraints are weaker (see Sec.3).} Nevertheless the NS$-$BH merger scenario is well testable in the era of GW astronomy \citep{Mandel2015}.  GRB 060614, if indeed from a NS$-$BH merger, is within the expected advanced LIGO/VIRGO sensitivity range \citep{Li2016a} and a further examination reveals a very promising detection prospect of NS$-$BH merger events \citep{Li2016b}. Though the GRB/GW events are expected to be still rare in the next decade, the macronova/GW events are expected to be much more frequent. The statistical study can be helpful in revealing the macronova nature and hence yield a better understanding of the GRB central engine.

\section*{Acknowledgments}
We thank the referee for the helpful comments and YZF appreciates Dr. K. Hotokezaka for the communications. This work was supported in part by 973 Programme of China under grants No. 2013CB83700 and No. 2014CB845800, National Natural Science Foundation of China under grants 11273063,  11433009 and 11525313, the Chinese Academy of Sciences via the External Cooperation Program of BIC (No. 114332KYSB20160007).
\\

\clearpage

\end{document}